\documentclass[aps,
prl,
reprint,
preprintnumbers,
superscriptaddress,
footinbib,
amsfonts,
amssymb,
amsmath
]{revtex4-1}
\usepackage{bm,latexsym,mathrsfs,enumerate}
\usepackage{upgreek}
\usepackage[table,x11names]{xcolor}
\usepackage[breaklinks=true,
unicode=true,
urlcolor = RoyalBlue4,
colorlinks = true,
citecolor = RoyalBlue4,
linkcolor = RoyalBlue4
]{hyperref}
\usepackage{graphicx}
\usepackage{pgffor}
\graphicspath{{./imgs/}}
\usepackage[cal=boondoxo,scr=rsfs]{mathalfa}
\renewcommand{\vec}[1]{\bm{#1}}
\usepackage[final]{pdfpages}
\makeatletter
\AtBeginDocument{\let\LS@rot\@undefined}
\makeatother

\begin{document}

\title{Curvilinear one-dimensional antiferromagnets}

\author{Oleksandr~V.~Pylypovskyi}
\thanks{These two authors contributed equally}
\affiliation{Helmholtz-Zentrum Dresden-Rossendorf e.V., Institute of Ion Beam Physics and Materials Research, 01328 Dresden, Germany}
\affiliation{Taras Shevchenko National University of Kyiv, 01601 Kyiv, Ukraine}

\author{Denys~Y.~Kononenko}
\thanks{These two authors contributed equally}
\affiliation{Taras Shevchenko National University of Kyiv, 01601 Kyiv, Ukraine}
\affiliation{Institute for Theoretical Solid State Physics, IFW Dresden, 01069 Dresden, Germany}

\author{Kostiantyn~V.~Yershov}
\affiliation{Institute for Theoretical Solid State Physics, IFW Dresden, 01069 Dresden, Germany}
\affiliation{Bogolyubov Institute for Theoretical Physics of National Academy of Sciences of Ukraine, 03143 Kyiv, Ukraine}

\author{Ulrich~K.~R{\"o}{\ss}ler}
\affiliation{Institute for Theoretical Solid State Physics, IFW Dresden, 01069 Dresden, Germany}

\author{Artem~V.~Tomilo}
\affiliation{Taras Shevchenko National University of Kyiv, 01601 Kyiv, Ukraine}

\author{J\"{u}rgen Fa\ss bender}
\affiliation{Helmholtz-Zentrum Dresden-Rossendorf e.V., Institute of Ion Beam Physics and Materials Research, 01328 Dresden, Germany}

\author{Jeroen~van~den~Brink}
\affiliation{Institute for Theoretical Solid State Physics, IFW Dresden, 01069 Dresden, Germany}
\affiliation{Institute for Theoretical Physics, TU Dresden, 01069 Dresden, Germany}

\author{Denys~Makarov}
\email{d.makarov@hzdr.de}
\affiliation{Helmholtz-Zentrum Dresden-Rossendorf e.V., Institute of Ion Beam Physics and Materials Research, 01328 Dresden, Germany}

\author{Denis~D.~Sheka}
\email{sheka@knu.ua}
\affiliation{Taras Shevchenko National University of Kyiv, 01601 Kyiv, Ukraine}
	
\date{July 22, 2020}

\begin{abstract}

Antiferromagnets host exotic quasiparticles, support high frequency excitations and are key enablers of the prospective spintronic and spin-orbitronic technologies. Here, we propose a concept of a curvilinear antiferromagnetism where material responses can be tailored by a geometrical curvature without the need to adjust material parameters. We show that an intrinsically achiral one-dimensional (1D) curvilinear antiferromagnet behaves as a chiral helimagnet with geometrically tunable Dzyaloshinskii--Moriya interaction (DMI) and orientation of the N\'{e}el vector. The curvature-induced DMI results in the hybridization of spin wave modes and enables a geometrically-driven local minimum of the low frequency branch. This positions curvilinear 1D antiferromagnets as a novel platform for the realization of geometrically tunable chiral antiferromagnets for antiferromagnetic spin-orbitronics and fundamental discoveries in the formation of coherent magnon condensates in the momentum space.

\end{abstract}

\maketitle

\textit{Introduction--} 
Antiferromagnets (AFMs) emerged as a versatile material science platform, which enabled numerous fundamental discoveries including observation of monopole quasiparticles in frustrated systems~\cite{Castelnovo08,Nagaosa12,Zvyagin13} and collective quantum effects, such as spin superfluidity~\cite{Takei14,Proukakis17,Baltz18} and Bose--Einstein condensation (BEC) of magnetic excitations ~\cite{Rice02,Giamarchi08,Proukakis17}. This trend is even further facilitated by the advent of antiferromagnetic spintronics~\cite{Gomonay17,Baltz18,Smejkal18,Manchon19} and related novel physical concepts of staggered spin-orbit torques~\cite{Zhang14,Zelezny14,Manchon19}. These effects are specific to AFMs possessing broken inversion symmetry in a local environment, which is also a source of Dzyaloshinskii--Moriya interaction (DMI)~\cite{Dzyaloshinsky57,Bogdanov89r,Qaiumzadeh18a}. The presence of DMI is peculiar for noncollinear AFMs characterized by weak ferromagnetism and chiral helimagnetism ~\cite{Bogdanov02a}. DMI significantly affects dynamics of solitary excitations in AFMs including much higher domain wall velocities~\cite{Qaiumzadeh18} and absence of the gyroforce (Magnus force) for skyrmions~\cite{Barker16,Shen19c}. The portfolio of material systems available for these studies is very limited due to the stringent requirement to the magnetic symmetry of chiral AFMs. This requirement renders the progress in AFM-related fundamental and technological research to depend on time consuming material screening and optimization of intrinsic chiral properties of AFMs.

For ferromagnets (FMs) chiral responses in nanowires and thin films can be tailored by using curvilinear geometries~\cite{Gaididei14,Sheka15c,Pylypovskyi16,Sheka20a}. This framework, known as curvilinear magnetism~\cite{Streubel16a,Fernandez17,Fischer20,Vedmedenko20}, allows to induce magnetochiral effects in otherwise conventional achiral FMs~\cite{Pylypovskyi16,Otalora16,Vojkovic17}. In contrast to FMs, no theory of curvilinear antiferromagnetism is available to date. Therefore, the appealing approaches to use geometrical curvatures to enable chiral properties in AFMs are not explored. If available, it would be possible to decouple the design of chiral responses of AFMs and their intrinsic magnetic properties. We emphasize that the case of curvilinear AFMs is fundamentally different to curvilinear FMs primarily due to the necessity to self-consistently account for the \emph{mutual} interplay of several (at least two for AFM \textit{vs} one for FM) fields of magnetization with the geometrical curvature. This is directly related to the physical nature of the order parameter, which is director in AFMs in contrast to the magnetization vector in FMs.

In this Letter, we put forth a fundamental foundation  of curvilinear one-dimensional (1D) AFMs. We explore curvature effects in a prototypical AFM system, namely, spin chain where spins are coupled via local exchange and long-range dipolar interactions. The nearest-neighbor exchange interaction brings about two geometry-induced responses: the intrinsically achiral curvilinear AFM spin chain behaves as a chiral helimagnet with a geometrically tunable DMI and a biaxial anisotropy. We show that generic curvilinear 1D AFMs exhibit the full set of Lifshitz invariants whose strength is determined by the local torsion and curvature.  We apply our theory to analyze static and dynamic responses of helical AFM spin chains to demonstrate consequences of the coupling between the geometry and the AFM order parameter. Spin chains arranged along space curves with non-zero torsion exhibit a magnetic phase transition from homogeneous to periodic states, which is tunable by controlling geometrical parameters. The appearance of the curvature-induced DMI results in the hybridization of spin wave modes in linear dynamics and opens a possibility to investigate coherent and long-living magnon state in the DMI-induced minimum of the dispersion curve.

\begin{figure*}[t]
	\includegraphics[width=\textwidth]{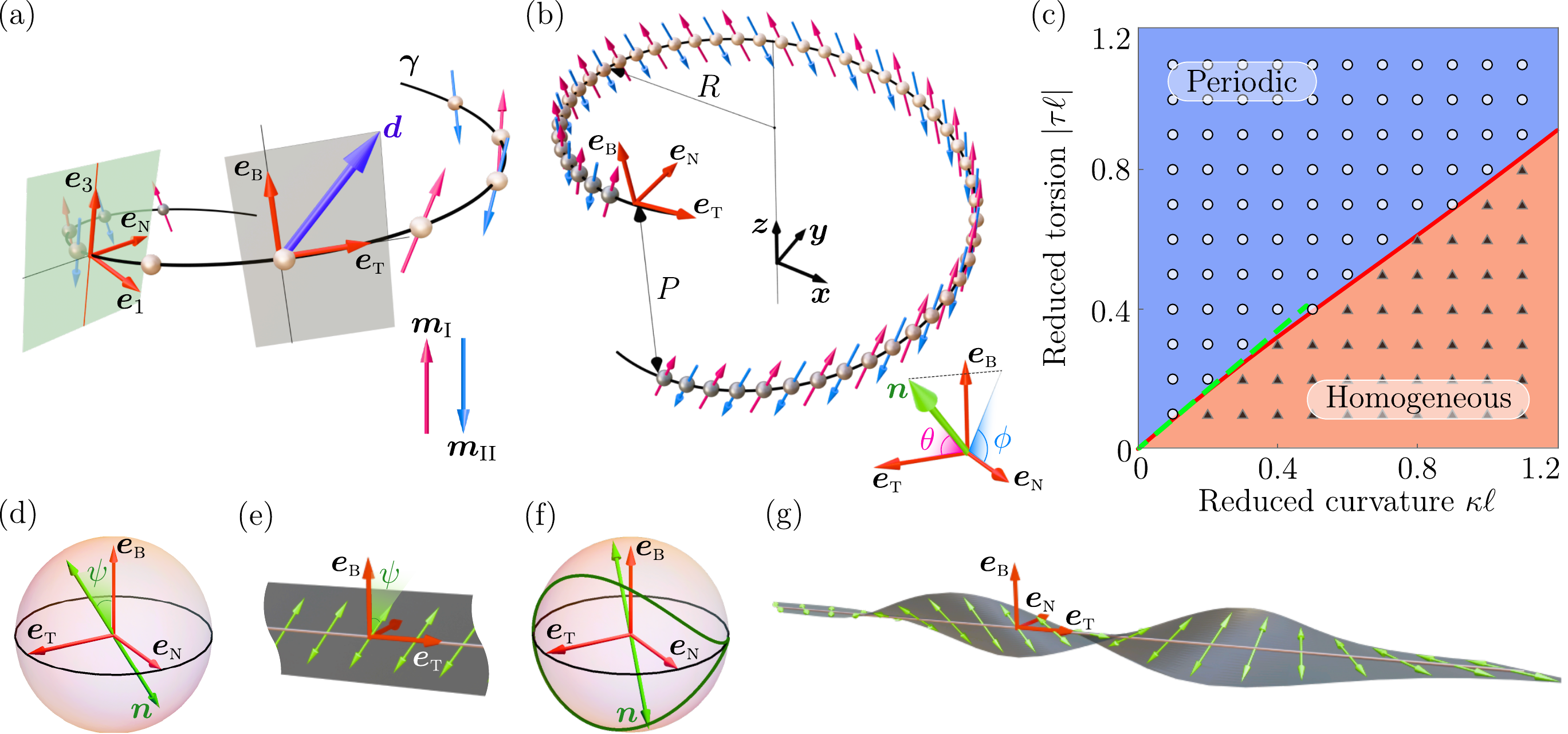}
	\caption{(Color online) (a) Schematics of the antiferromagnetic spin chain $\vec{\gamma}$. Magnetic sublattices with magnetization $\vec{m}_\textsc{i}$ and $\vec{m}_\textsc{ii}$ are shown by magenta and light-blue arrows. The Dzyaloshinskii vector $\vec{d}$ (dark-blue) lies in the TB plane given by the TNB basis $\vec{e}_\textsc{t,n,b}$. Hard and easy anisotropy axes are labeled by $\vec{e}_1$ and $\vec{e}_3$, respectively. (b) Helix spin chain with radius $R$ and pitch $P$.  The AFM order parameter (N\'{e}el vector) $\vec{n}$ parameterized by angles $\theta$ and $\phi$ is shown by the green arrow. (c)~Diagram of equilibrium states for a helix spin chain. Open symbols and triangles correspond to periodic and homogeneous states, respectively, obtained in spin-lattice simulations. Solid red curve shows the boundary between the states. Dashed green line shows the asymptotic of the boundary $\tau_b\approx0.85\kappa$ for $\kappa\ell\ll1$.	Schematics of the homogeneous (d,e) and periodic states (f,g) in the TNB reference frame. Bloch spheres (d,f) illustrate the trajectories of $\vec{n}$. The tilt angle $\psi \approx \ell^2\kappa\tau$. 
	} 
	\label{fig:helix_statics}
\end{figure*}

\textit{Model of a curvilinear AFM--} 
We start with a classical spin chain taking into account the AFM nearest-neighbor exchange and dipolar interaction. Its static and dynamic properties are determined by the Landau--Lifshitz equation $\hslash S \mathrm{d} \vec{m}_i/\mathrm{d}t  = \vec{m}_i \times \partial \mathscr{H}/\partial \vec{m}_i$ with the Hamiltonian specific to the collinear intrinsically achiral AFM
\begin{equation}\label{eq:Hamiltonian}
\mathscr{H} = -\dfrac{J S^2}{2} \sum_i \vec{m}_i\cdot \vec{m}_{i+1} - \dfrac{\mu}{2} \sum_i \vec{m}_i\cdot \vec{H}_i^d.
\end{equation}
Here, $\vec{m}_i$ is the unit magnetic moment of $i$-th site, $\hslash$ is the Planck constant, $S$ is the spin length, $J < 0$ is the exchange integral, $\mu = g\mu_\textsc{b}S$ is the total magnetic moment of one site with $g$ being Land\'{e} factor, and $\mu_\textsc{b}$ being Bohr magneton. The dipolar field at $i$-th site reads $\vec{H}_i^d = -\mu \sum_{ j = 1+i }^\infty [ \vec{m}_j r_{ij}^2 - 3\vec{r}_{ij}(\vec{m}_j\cdot \vec{r}_{ij})]/r_{ij}^5 $ with $\vec{r}_{ij}$ being the radius-vector between $i$-th and $j$-th sites and the distance between neighboring sites equal $a$. We assume that the positions of all magnetic sites are described by a space curve $\vec{\gamma}(s)$ with $s$ being the arc-length characterized by the curvature $\kappa(s)$ and torsion $\tau(s)$. The local reference frame can be chosen as the Frenet--Serret frame with tangential, normal and binormal vectors $\vec{e}_\textsc{t,n,b}$, see Fig.~\ref{fig:helix_statics}(a).

The continuum counterpart of the spin-lattice model is formulated based on two vector fields, namely the total magnetization $\vec{m}(s) = (\vec{m}_\textsc{i} + \vec{m}_\textsc{ii})/2$ and N\'{e}el vector $\vec{n}(s) = (\vec{m}_\textsc{i} - \vec{m}_\textsc{ii})/2$. The fields $\vec{m}_{\textsc{i},\textsc{ii}} = \vec{m}_{\textsc{i},\textsc{ii}}(s)$ correspond to the two sublattices of the AFM.  In the long-wave approximation, the density of Lagrangian $L = \int\mathscr{L} \mathrm dx$, corresponding to the curvilinear AFM reads
\begin{subequations} \label{eq:Lagrangian-curv}
\begin{equation} \label{eq:Lagrangian-curv-L}
\mathscr{L} = \dfrac{M_\textsc{s}^2}{\gamma_0^2\Lambda}\dot{\vec{n}}^2 - \mathscr{E}
\end{equation}	
with the overdot corresponding to the derivative with respect to time. The effective energy density $\mathscr{E}$ is written as
\begin{equation} \label{eq:Lagrangian-curv-E}
\begin{aligned}
\mathscr{E} = & \mathscr{E}_\text{x} + \mathscr{E}_\textsc{dm} + \mathscr{E}_\textsc{a} + K(\vec{n}\cdot \vec{e}_\textsc{t})^2\\
\mathscr{E}_\text{x} = & A n_\alpha' n_\alpha', \quad \mathscr{E}_\textsc{dm} = A\mathscr{F}_{\alpha \beta}(n_\alpha n_\beta'  - n_\beta n_\alpha'), \\
\mathscr{E}_\textsc{a} = &A\mathscr{K}_{\alpha\beta} n_\alpha n_\beta,\quad \alpha, \beta \in \text{T,N,B},
\end{aligned}
\end{equation}	
\end{subequations}
where $M_\textsc{s} = \mu/(2a)$ is the magnetization of one sublattice, $\gamma_0$ is the gyromagnetic ratio, $\Lambda = 2|J|S^2/a$ is the constant of the uniform exchange, $A = |J|S^2a / 2$ is the exchange stiffness, and $K\approx 2.7 \mu^2/a^4$ is the hard axis anisotropy constant induced by the dipolar interaction, see Supplemental Material \footnote{See Supplemental Material at \href{http://ritm.knu.ua}{Link provided by the publisher} for details of analytical calculations and simulations, which includes Refs.~\cite{Turov01en,Baryakhtar79,Andreev80, Streubel16a,Vedmedenko20, Sheka15,NIST10,Slastikov12, Kronmueller07v2,Sheka15c, Waldmann03,Furukawa08, Ivanov95e,SLaSi, Pylypovskyi13f,Weisstein03,unicc}}. The model~\eqref{eq:Lagrangian-curv} is valid for $K \ll \Lambda$ and the space curve $\vec{\gamma}$ possessing consequent turns separated by a distance significantly larger than the lattice constant $a$. In this approximation, $|\vec{m}|\ll |\vec{n}|$ and $\vec{n}$ can be considered as a unit director. In \eqref{eq:Lagrangian-curv-E} the Einstein summation rule is applied and prime means derivative with respect to $s$. The Frenet tensor $\mathscr{F}_{\alpha\beta}$ has four nonzero components $\mathscr{F}_\textsc{tn} = -\mathscr{F}_\textsc{nt} = \kappa$ and $\mathscr{F}_\textsc{nb} = -\mathscr{F}_\textsc{bn} = \tau$. The characteristic length and time scales are given by the magnetic length $\ell=\sqrt{A/K}$ and the frequency of the AFM resonance $\omega_0 = c/\ell$ with $c=\gamma_0\sqrt{\Lambda A}/M_s$ being the characteristic magnon speed.
The exchange energy density expands into three terms, with only one, $\mathscr{E}_\text{x}$, possessing the form of a regular inhomogeneous exchange in straight spin chains. 

The term $\mathscr{E}_\textsc{dm}$ can be written as the functional form of a DMI, $\mathscr{E}_\textsc{dm} = \vec{d} \cdot [ \vec{n} \times \vec{n}' ] $. This term is allowed in crystals with magnetic symmetry groups $C_n$ and $S_4$ acting on 1D magnetic textures~\cite{Bogdanov89r}. However, its origin is not the spin-orbit interaction as for the case of intrinsic DMI but the exchange interaction. The vector $\vec{d} = d_\textsc{t} \vec{e}_\textsc{t} + d_\textsc{b} \vec{e}_\textsc{b}$ acts as the Dzyaloshinskii vector with components  $d_\textsc{t} = 2A\tau$ and $d_\textsc{b} = 2A\kappa$.  This DMI corresponds to the full set of Lifshitz invariants, allowed in a 1D magnet. The DMI vector $\vec{d}$ is linear with respect to $\tau$ and $\kappa$, which allows strong chiral effects in curvilinear 1D AFMs. The strength of the curvature-induced DMI can be estimated as the relation to the exchange stiffness. For instance, in the case of a Mn-DNA chain (A-DNA form) bent to the radius of 15\,nm, the $ad_\textsc{t,b}/A$ is about 0.05~\footnote{Magnetic parameters of Mn-DNA $S = 5/2$, $a = 0.344$\,nm and $J = 9.6\times 10^{-25}$\,J are taken from~\cite{Mizoguchi05}.}. This value is comparable with the intrinsic chiral properties of KMnF$_3$ used for the discussion of dynamics of 1D solitons~\cite{Qaiumzadeh18,Pan18} ($aD/A = 0.036 $ with $D$ being the constant of the non-uniform DMI), where ultrafast motion of AFM domain walls was predicted~\cite{Qaiumzadeh18}.

In addition to the linear in $\tau$ and $\kappa$ DMI terms,  the expression for energy density $\mathscr{E}$ contains  weaker bilinear terms, representing a curvature-induced anisotropy $\mathscr{E}_\textsc{a}$ whose coefficients are given by the tensor $\mathscr{K}_{\alpha\beta} = \mathscr{F}_{\alpha\gamma}\mathscr{F}_{\beta\gamma}\propto \kappa^2, \tau^2$, $\kappa \tau$. It contains non-diagonal terms, causing the tilt of $\vec{n}$ within the rectifying surface formed by $\vec{e}_\textsc{t}$ and $\vec{e}_\textsc{b}$. The presence of the two anisotropies (hard axis stemming from the dipolar interaction and easy axis stemming from the exchange interaction) renders a curvilinear AFM spin chain to behave as a biaxial AFM. The directions of the primary hard axis $\vec{e}_1$ and secondary easy axis $\vec{e}_3$ are determined by the diagonalization of the tensor of the total anisotropy $A\mathscr{K}_{\alpha\beta}+\delta_{1\alpha}\delta_{1\beta}K$ with $\delta_{\alpha\beta}$ being Kronecker delta, see Fig.~\ref{fig:helix_statics}(a). The axis $\vec{e}_1$ lies within the rectifying surface. The anisotropy induced by the dipolar interaction is the strongest one and defines the plane, where the N\'{e}el vector rotates. The direction of the vector $\vec{n}$ within the easy plane is given by the curvature-induced anisotropy $\mathscr{E}_\textsc{a}$. The system has no competing easy axis anisotropy terms. This means that independent of the strength of $\mathscr{E}_\textsc{a}$ it govern the orientation of the N\'{e}el vector even for $A\kappa^2, A\tau^2 \ll K$.

As a result, a generic curvilinear achiral 1D AFM will behave as a chiral helimagnet with the DMI strength and the orientation of the N\'{e}el vector determined by the geometrical parameters, i.e., curvature and torsion.

\textit{Ground state of AFM helix chains--}
To illustrate the behavior of curvilinear AFM spin chains, described by~\eqref{eq:Lagrangian-curv}, we analyze a helix chain as the prototypical curvilinear systems possessing a constant curvature and torsion. The geometry of a helix is characterized by the radius $R = \kappa/(\kappa^2+\tau^2)$ and pitch $P = 2\pi \tau / (\kappa^2 + \tau^2)$, see Fig.~\ref{fig:helix_statics}(b). It is convenient to introduce the angular parametrization of the N\'{e}el vector $\vec{n} = \vec{e}_\textsc{t} \cos\theta + \vec{e}_\textsc{n}  \sin\theta\cos\phi + \vec{e}_\textsc{b}  \sin\theta \sin\phi$ with $\theta=\theta(s,t)$ and  $\phi=\phi(s,t)$ being polar and azimuthal angles, respectively. Taking into account the N\'{e}el vector is a director, states with $(\theta, \phi)$ and $(\pi - \theta, \phi \pm \pi)$ are equivalent. Then, the linear energy density reads
\begin{equation} \label{eq:helix-energy-density}
\begin{aligned}
\mathscr{E} & = A(\theta' + \kappa \cos\phi)^2 \\
& + A \left[ \sin\theta(\phi' + \tau) - \kappa\cos\theta\sin\phi \right]^2 + K \cos^2\theta.
\end{aligned}
\end{equation}
As a biaxial chiral helimagnet, helix spin chains support homogeneous and periodic equilibrium states dependent on the strength of the DMI, see Fig.~\ref{fig:helix_statics}(c). For the case of the homogeneous state, which is realized for $\tau < \tau_b(\kappa) \approx 0.85 \kappa$ at $\kappa \ell \ll 1$, see Fig.~\ref{fig:helix_statics}(d,e) and Supplemental Material \cite{Note1}, the orientation of the N\'{e}el vector is given by $\theta_\text{hom} = \pi/2 - \psi$ and $\phi_\text{hom} = \pi/2$, where $ \psi \approx \ell^2 \kappa\tau$ and $\kappa\ell,|\tau|\ell\ll 1$. 

The periodic state can be stabilized in systems possessing torsion $\tau > \tau_b(\kappa)$. In the periodic state, the N\'{e}el vector is almost  uniform in the plane perpendicular to the helix axis and  modulated in the local reference frame, see Fig.~\ref{fig:helix_statics}(f,g). The emergence of the periodic state is a consequence of the exchange-induced DMI,~$\mathscr{E}_\textsc{dm}$, with the main contribution given by the torsion-related term $d_\textsc{t}$. When the curvature is much smaller than the torsion, the state can be described as the Dzyaloshinskii spiral~\cite{Dzyaloshinski65} with $\theta_\text{per} = \pi/2$ and $\phi_\text{per} = - \tau s$. The boundary between the homogeneous and periodic states $\tau_b(\kappa)$ is plotted by the solid red line in Fig.~\ref{fig:helix_statics}(c).

\begin{figure}[t]
\includegraphics[width=\columnwidth]{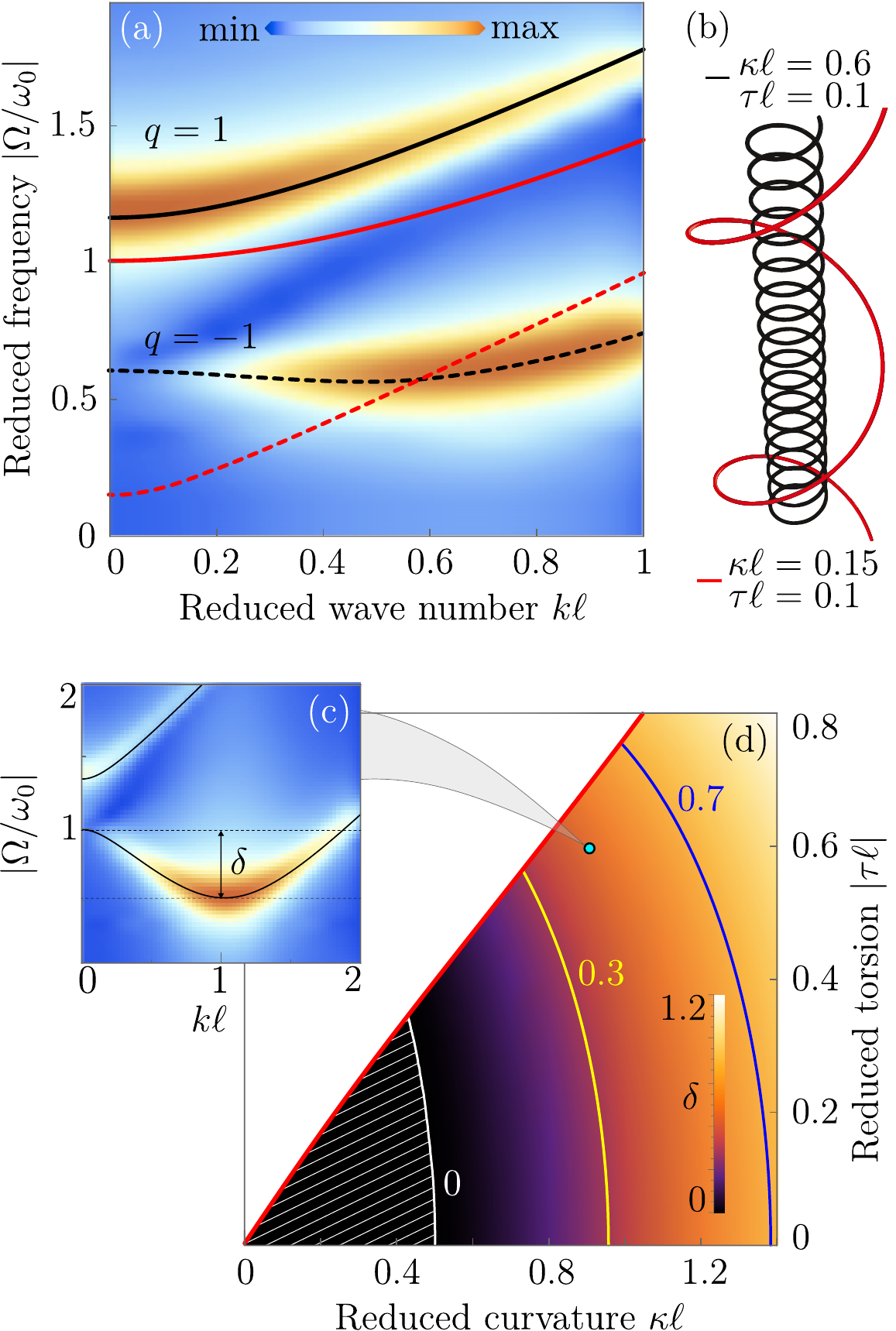}
\caption{(Color online) (a) Spin-wave dispersion~\eqref{eq:eigenfrequency} for helical AFM spin chains with $\tau\ell = 0.1$ and two curvatures $\kappa\ell = 0.6$ (black) and $\kappa\ell = 0.15$ (red). The result of spin-lattice simulations is shown by the background color for a helical spin chain with the geometry with $\tau\ell = 0.1$ and $\kappa\ell = 0.6$. (b) Helix geometries calculated in (a). (c) Spin-wave dispersion~\eqref{eq:eigenfrequency} and simulations for $\kappa\ell = 0.9$ and $\tau\ell = 0.6$. The depth of the minimum in the acoustic branch is shown by $\delta$. (d) The depth $\delta$ for different curvatures and torsions within the homogeneous ground state (below red line, same as in Fig.~\ref{fig:helix_statics}(c)). Dashed line corresponds to the absence of minimum.}
\label{fig:helix_spectrum} 
\end{figure}

It is instructive to compare the results above with FM spin chains, where dipolar interactions induce \emph{easy axis} anisotropy. In contrast to a FM helices \cite{Sheka15c}, the phase transition between the periodic and homogeneous states in AFMs has no threshold in curvature. Hence, the transition to the periodic state in the case of AFM helical chains can be observed for very small curvatures. This is a consequence of the specificity of the curvilinear AFM systems where the stability of the state is given by the weak easy axis anisotropy stemming from the exchange interaction. Therefore, effects of curvilinearity in AFMs are much stronger than in FMs.

\textit{Linear dynamics--}
To describe linear excitations in a curved AFM helix chain, we consider the homogeneous magnetic state. The Euler--Lagrange equations for the Lagrangian~\eqref{eq:Lagrangian-curv} are linearized by $\theta(s, t) = \theta_\text{hom} + \vartheta(s, t)$ and $\phi(s, t) = \phi_\text{hom} + \varphi(s, t) / \sin \theta_\text{hom}$. Here, $\vartheta(s, t)$ and $\varphi(s, t)$ are small deviations from the equilibrium state. The corresponding equations 
read
\begin{align} \label{eq:spectrum-motion} 
\begin{split}
A \vartheta'' - A c^{-2} \ddot{\vartheta} & = K_0\vartheta + D_3 \varphi',
\\
A \varphi'' - A c^{-2} \ddot{\varphi}     & = K_3\varphi   -D_3 \vartheta',
\end{split}
\end{align}
where $K_{0,3}$ and $D_3$ are functions of curvature, acting as the effective anisotropy and DMI coefficients, respectively, see Supplemental Material~\cite{Note1}. For a large curvature radius and small torsion, $K_0 \approx A\left(\ell^{-2}+\kappa^2-\tau^2 \right)$, $K_3 \approx A\kappa^2$ and $D_3 \approx 2A \kappa$. 
The dispersion law can be written using the substitution of plane waves $\vartheta(s,t) = \vartheta_k \cos(k s - \Omega t)$ and $\varphi(s,t) = \varphi_k \sin(k s - \Omega t)$, where $\vartheta_k$ and $\varphi_k$ are small amplitudes, $k$ is the wave number, and $\Omega$ is frequency. The dispersion reads
\begin{equation} \label{eq:eigenfrequency}
\frac{\Omega^2}{c^2} = k^2 + \frac{K_0+K_3}{2A} + \frac{q}{2A} \sqrt{(K_0-K_3)^2 + 4D_3^2 k^2}.
\end{equation}
We note that the dispersion curve is similar to flat biaxial AFMs with DMI~\cite{Qaiumzadeh18} and remains symmetric with respect to the sign of the momentum $k$. Yet, the geometrical tunability of the anisotropy and DMI allows to unveil new physics of collective excitations in curvilinear 1D AFMs.

The spin-wave spectrum~\eqref{eq:eigenfrequency} superimposed with spin-lattice simulations~\cite{Note1} is shown in Fig.~\ref{fig:helix_spectrum}(a) for two helix geometries [Fig.~\ref{fig:helix_spectrum}(b)]. The high frequency optical branch with $q=1$ is always gapped and the change of the geometry affects only the gap due to the curvature-induced anisotropy, $\Omega_{q=+1}^\text{gap} = \omega_0 + c\ell(\kappa^2-\tau^2)/2$. In contrast, there is a strong qualitative impact of the curvature on the low frequency branch. While it is gapless for a straight spin chain~\cite{Ivanov95e}, the gap  $\Omega_{q=-1}^\text{gap}\approx c\kappa$ appears for any finite curvature as a results of the the spin-wave hybridization, forming a low-frequency optical branch with $q=-1$, see Fig.~\ref{fig:helix_spectrum}(a).
The curvature-induced DMI results in the emergence of a region with a negative group velocity followed by a local minimum at $k = k_\text{min}$ on the dispersion curve with the depth $\delta$, see Fig.~\ref{fig:helix_spectrum}(a) and (c). The presence of a negative group velocity is also observed for multiferroics~\cite{Sousa08} and exchange-dipolar modes in AFM thin films~\cite{Stamps84,Stamps87}. The depth of the minimum increases with $\kappa$ and $\tau$, see Fig.~\ref{fig:helix_spectrum}(d). 
The possibility to realize magnon ground states not in equilibrium ($k \neq 0$ at minimum energy)~\cite{Demokritov13} renders curvilinear 1D AFMs a flexible platform to study coherent excitations for spin superfluidity~\cite{Yuan18b,Sonin19,Evers20} and BEC of magnons~\cite{Demokritov06,Serga14,Clausen15,Bozhko16} with taking into account a proper pumping and magnon thermodynamics.

\textit{Conclusions--} 
We develop a theory of curvilinear one-dimensional antiferromagnets. We demonstrate that the intrinsically achiral curvilinear AFM spin chain behaves as a biaxial chiral helimagnet with geometrically tunable DMI and anisotropy. The curvature-induced DMI results in the hybridization of magnon modes in the chain. The low frequency branch possesses a local minimum supporting a long-living magnon state, which allows to consider 1D curvilinear AFMs as the platform for the realization of BEC of magnons in $k$-space. Furthermore, the symmetry and strength of the geometry-induced DMI opens perspectives for applications in antiferromagnetic spin-orbitronics, e.g. for ultrafast dynamics of chiral domain walls~\cite{Qaiumzadeh18,Pan18}. We consider copper-based~\cite{Zhang16e} and DNA-based metal-organic frameworks~\cite{Zhang05,Yamaguchi05a,Mizoguchi07,Samanta13} as a promising materials for experimental validation of our predictions. For instance, one can expect the strength of curvature-induced DMI as $ad_\textsc{t,b}/A \approx 0.05$ which is comparable with AFMs supporting chiral domain walls.

\textit{Acknowledgments.--}
This paper is dedicated to the memory of the wonderful physicist Yuri Gaidiei, who recently passed away. We thank  U. Nitzsche for technical support. D.Y.K. and K.V.Y. acknowledge financial support from UKRATOP-project~(funded by BMBF under reference 01DK18002). In part, this work was supported by the Program of Fundamental Research of the Department of Physics and Astronomy of the National Academy of Sciences of Ukraine~(Project No.~0116U003192), by the Alexander von Humboldt Foundation (Research Group Linkage Programme), DFG MA 5144/22-1, and by Taras Shevchenko National University of Kyiv (Project No. 19BF052-01).

%
%

%



\section*{Supplementary material}

# Supplemental Material to
# "Curvilinear one-dimensional antiferromagnets"

Oleksandr V. Pylypovskyi,[1,2,*] Denys Y. Kononenko,[2,3,*] Kostiantyn V. Yershov,[3,4] Ulrich K. Rößler,[3] Artem V. Tomilo,[2] Jürgen Faßbender,[1] Jeroen van den Brink,[3,5] Denys Makarov,[1,†] and Denis D. Sheka[2,‡]

[1]*Helmholtz-Zentrum Dresden-Rossendorf e.V., Institute of Ion Beam Physics and Materials Research, 01328 Dresden, Germany*
[2]*Taras Shevchenko National University of Kyiv, 01601 Kyiv, Ukraine*
[3]*Institute for Theoretical Solid State Physics, IFW Dresden, 01069 Dresden, Germany*
[4]*Bogolyubov Institute for Theoretical Physics of National Academy of Sciences of Ukraine, 03143 Kyiv, Ukraine*
[5]*Institute for Theoretical Physics, TU Dresden, 01069 Dresden, Germany*

(Dated: July 22, 2020)

The supplementary information provides details on analytical calculations and numerical simulations.

## I. THE GEOMETRY

To describe magnetic properties of a curved antiferromagnetic (AFM) chain we choose a curvilinear reference frame adapted to the geometry of the object. Let $\boldsymbol{\gamma}(s)$ be a one-dimensional (1D) curve embedded in the three-dimensional (3D) space $\mathbb{R}^3$ with $s$ being the arc length. The local TNB reference frame reads

$$\boldsymbol{e}_{\text{T}} = \boldsymbol{\gamma}', \qquad \boldsymbol{e}_{\text{N}} = \frac{\boldsymbol{e}'_{\text{T}}}{|\boldsymbol{e}'_{\text{T}}|}, \qquad \boldsymbol{e}_{\text{B}} = \boldsymbol{e}_{\text{T}} \times \boldsymbol{e}_{\text{N}} \tag{S1}$$

with prime denoting the derivative with respect to $s$, vectors $\boldsymbol{e}_{\text{T}}$, $\boldsymbol{e}_{\text{N}}$ and $\boldsymbol{e}_{\text{B}}$ being tangential, normal and binormal basis vectors, respectively. Their differential properties are given by the Frenet–Serret formulae:

$$\boldsymbol{e}'_\alpha = \mathscr{F}_{\alpha\beta}\boldsymbol{e}_\beta, \qquad \|\mathscr{F}_{\alpha\beta}\| = \begin{pmatrix} 0 & \kappa & 0 \\ -\kappa & 0 & \tau \\ 0 & -\tau & 0 \end{pmatrix}, \tag{S2}$$

where $\kappa$ is the curvature, $\tau$ is the torsion of the curve $\boldsymbol{\gamma}$ and Greek indices $\alpha, \beta \in \text{T,N,B}$.

The geometry of a helix with the radius $R$ and pitch $P$ reads

$$\boldsymbol{\gamma} = \hat{\boldsymbol{x}} R \cos \frac{s}{s_0} + \hat{\boldsymbol{y}} R \sin \frac{s}{s_0} + \hat{\boldsymbol{z}} \frac{CPs}{2\pi s_0}, \tag{S3}$$

where $C = \pm 1$ is the helix chirality and $s_0 = \sqrt{R^2 + P^2/(2\pi)^2}$. Here, the curvature and torsion read $\kappa = R/s_0^2$ and $\tau = CP/(2\pi s_0^2)$, respectively. The parametrization of a ring geometry can be obtained from Eq. (S3) using $\tau = 0$ and the periodic boundary condition $\boldsymbol{\gamma}(s) = \boldsymbol{\gamma}(s + 2\pi/\kappa)$.

## II. THE MODEL OF CURVILINEAR AFM

We consider an isotropic, intrinsically achiral chain of unit magnetic moments $\boldsymbol{m}_i$ labeled by index $i \in \mathbb{Z}$. The total magnetic moment of the site is $\mu = g\mu_{\text{B}}S$ with $g$ being the Landé factor, $\mu_{\text{B}}$ being the Bohr magneton, and $S$ being the spin length. In a continuum limit, the magnetic structure of a spin chain with the AFM nearest neighbors exchange coupling corresponds to a two-sublattice AFM with magnetization fields $\boldsymbol{m}_{\text{I,II}}$ of each sublattice. It is convenient to introduce the Néel director $\boldsymbol{n} = (\boldsymbol{m}_{\text{I}} - \boldsymbol{m}_{\text{II}})/2$ and the total magnetization vector $\boldsymbol{m} = (\boldsymbol{m}_{\text{I}} + \boldsymbol{m}_{\text{II}})/2$ for further analysis. Dynamics of $\boldsymbol{m}$ and $\boldsymbol{n}$ are described by the Landau–Lifshitz equations in the form [1]

$$\dot{\boldsymbol{m}} = \gamma_0 \left( \boldsymbol{n} \times \frac{\delta E}{\delta \boldsymbol{n}} + \boldsymbol{m} \times \frac{\delta E}{\delta \boldsymbol{m}} \right), \qquad \dot{\boldsymbol{n}} = \gamma_0 \left( \boldsymbol{n} \times \frac{\delta E}{\delta \boldsymbol{m}} + \boldsymbol{m} \times \frac{\delta E}{\delta \boldsymbol{n}} \right), \tag{S4}$$

---

* These two authors contributed equally
† d.makarov@hzdr.de
‡ sheka@knu.ua

where overdot means derivative with respect to time $t$, $\gamma_0$ is the gyromagnetic ratio and $E = \int \mathscr{E} \mathrm{d}s$ is the macroscopic counterpart of the Hamiltonian (1) in the main text. The energy density reads

$$\mathscr{E} = \Lambda \boldsymbol{m}^2 + A(\boldsymbol{n}')^2 + K(\boldsymbol{n} \cdot \boldsymbol{e}_{\mathrm{T}})^2 - K_m (\boldsymbol{m} \cdot \boldsymbol{e}_{\mathrm{T}})^2, \tag{S5}$$

where $\Lambda = 2|J|S^2/a$ is the constant of the uniform exchange, $A = 2|J|S^2 a/2$ is the exchange stiffness with $J < 0$ being the exchange constant, and $a$ being the distance between neighboring magnetic moments. The quantities $K$ and $K_m$ are anisotropy constants. In this work, their microscopic origin is the dipole interaction, see Sec. III below.

Assuming that $|K|, |K_m| \ll \Lambda$ and using a long wave approximation, the total magnetization and its derivatives are small, then $|\boldsymbol{m}|, |\boldsymbol{m}'|, |\boldsymbol{m}''| \ll |\boldsymbol{n}|$. Therefore, $\boldsymbol{n}$ can be considered as a unit vector. Then, Eqs. (S4) can be reduced to the equation of motion for the Néel vector as [2, 3]

$$\boldsymbol{n} \times \left( c^2 \boldsymbol{n}'' - \ddot{\boldsymbol{n}} + \omega_0^2 n_{\mathrm{T}} \boldsymbol{e}_{\mathrm{T}} \right) = 0, \tag{S6}$$

where $c = \gamma_0 \sqrt{\Lambda A}/M_{\mathrm{s}}$ is the characteristic magnon velocity and $\omega_0 = \gamma_0 \sqrt{\Lambda K}/M_{\mathrm{s}}$ is the characteristic magnon frequency with $M_{\mathrm{s}} = \mu/(2a)$. Eq. (S6) can be also obtained within the Lagrange formalism [1]. The reduced Lagrangian density of the 1D AFM reads

$$\mathscr{L} = \frac{\mathscr{L}}{K} = (\partial_\tau \boldsymbol{n})^2 - \mathscr{E}, \qquad \mathscr{E} = \frac{\mathscr{E}}{K} = (\partial_\xi \boldsymbol{n})^2 + (\boldsymbol{n} \cdot \boldsymbol{e}_{\mathrm{T}})^2, \tag{S7a}$$

where $\tau = \omega_0 t$ is the reduced time and $\xi = s/\ell$ is the reduced coordinate.

Introducing curvature and torsion by bending the magnetic chain, the properties of magnetic states of the chain can be drastically changed. The dominant reason is hereby given by effective magnetic interactions caused by locally curved geometries: (i) curvilinear geometry-induced effective anisotropy and (ii) curvilinear geometry-induced effective Dzyaloshinskii-Moriya interaction (DMI) [4, 5]. These effective interactions are driven by the common isotropic exchange interaction in a curvilinear reference frame, which follows the chain. The transition into the curvilinear frame of reference is not a mathematical trick, but is physically conditioned by the presence of the magnetic interactions determined by the sample geometry. In particular, such geometry-dependent interaction in the current study is the effective anisotropy, originated from the dipolar interaction, see Sec. III: the direction of the anisotropy, $\boldsymbol{e}_{\mathrm{T}}$, is locally controlled by the local curvature and torsion of the AFM chain.

The energy of the AFM in the curvilinear reference frame (S1) can be calculated using the procedure, developed earlier to treat curvilinear ferromagnets (FMs) in Ref. [6], cf. Eq.(2b) of the main text:

$$\mathscr{E} = \mathscr{E}_{\mathrm{x}} + \mathscr{E}_{\mathrm{DM}} + \mathscr{E}_{\mathrm{A}} + (\boldsymbol{n} \cdot \boldsymbol{e}_{\mathrm{T}})^2, \qquad \mathscr{E}_{\mathrm{x}} = (\partial_\xi n_\alpha)(\partial_\xi n_\alpha), \qquad \mathscr{E}_{\mathrm{DM}} = \mathscr{F}_{\alpha\beta}(n_\alpha \partial_\xi n_\beta - n_\beta \partial_\xi n_\alpha), \qquad \mathscr{E}_{\mathrm{A}} = \mathscr{K}_{\alpha\beta} n_\alpha n_\beta. \tag{S7b}$$

Here, $\mathscr{F}_{\alpha\beta} = \ell \mathscr{F}_{\alpha\beta}$ is the reduced Frenet tensor, and $\mathscr{K}_{\alpha\beta} = \mathscr{F}_{\alpha\nu} \mathscr{F}_{\beta\nu}$ is the curvilinear geometry induced exchange driven anisotropy tensor. Its explicit form reads

$$\|\mathscr{K}_{\alpha\beta}\| = \begin{pmatrix} \varkappa^2 & 0 & -\varkappa\sigma \\ 0 & \varkappa^2 + \sigma^2 & 0 \\ -\varkappa\sigma & 0 & \sigma^2 \end{pmatrix}, \tag{S7c}$$

where $\varkappa = \kappa\ell$ is a reduced curvature and $\sigma = \tau\ell$ is a reduced torsion. It is useful to incorporate the constraint $|\boldsymbol{n}| = 1$ by means of the angular parametrization $\boldsymbol{n} = \boldsymbol{e}_{\mathrm{T}} \cos\theta + \boldsymbol{e}_{\mathrm{N}} \sin\theta \cos\phi + \boldsymbol{e}_{\mathrm{B}} \sin\theta \sin\phi$. Then the density of Lagrangian (S7) reads

$$\mathscr{L} = (\partial_\tau \theta)^2 + \sin^2\theta (\partial_\tau \phi)^2 - \mathscr{E}, \quad \mathscr{E} = (\partial_\xi \theta + \varkappa \cos\phi)^2 + [\sin\theta (\partial_\xi \phi + \sigma) - \varkappa \cos\theta \sin\phi]^2 + \cos^2\theta. \tag{S8}$$

### III. DIPOLAR INTERACTION IN SPIN CHAINS AS AN EFFECTIVE ANISOTROPY

Here, we will discuss the influence of the magnetic dipolar interaction on the anisotropic properties of an 1D AFM system. We start from a straight chain with spin sites placed along the $x$ axis. The energy of the dipolar interaction reads

$$\mathscr{H}_{\mathrm{D}} = \mu^2 \sum_{\substack{i=-\infty \\ j=1}}^{\infty} \left[ \frac{(\boldsymbol{m}_i \cdot \boldsymbol{m}_{i+j})}{r_{i,i+j}^3} - 3 \frac{(\boldsymbol{m}_i \cdot \boldsymbol{r}_{i,i+j})(\boldsymbol{m}_{i+j} \cdot \boldsymbol{r}_{i,i+j})}{r_{i,i+j}^5} \right], \tag{S9}$$

where $\boldsymbol{r}_{i,i+j} = \boldsymbol{r}_{i+j} - \boldsymbol{r}_i$ is a radius-vector between $i$-th and $(i+j)$-th site of the chain with $r_{i,i+j} = |\boldsymbol{r}_{i,i+j}|$. The first term has a rotational symmetry whereas the second one is responsible for anisotropic properties in the chain. Let us examine the influence of the second term in Eq. (S9) on a magnetic texture of the intrinsically isotropic spin chain.

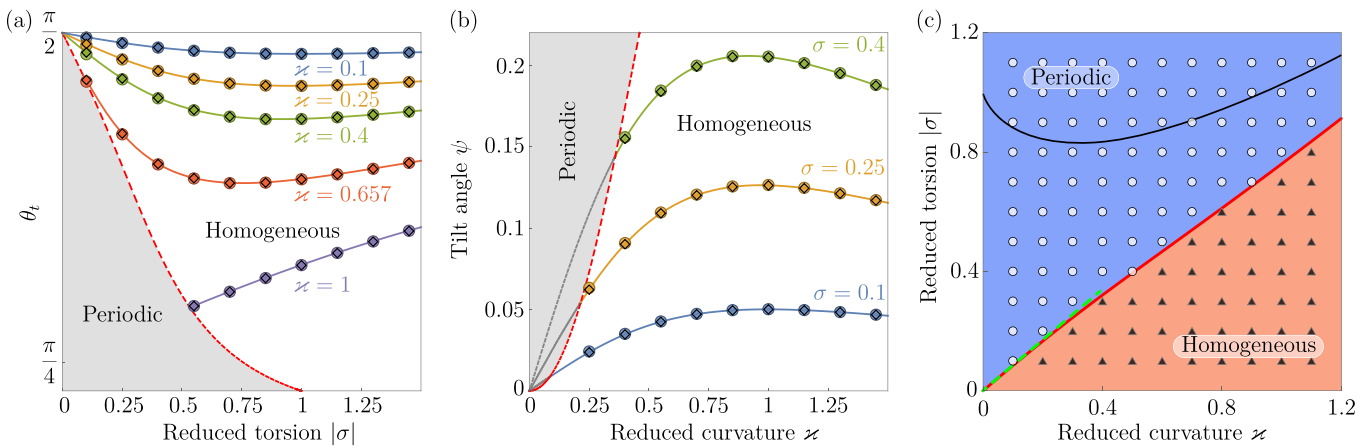

FIG. S1. (Color online) **Equilibrium states in FM and AFM helix chains**. (a) Homogeneous equilibrium state in FM helices with different curvature and torsion; solid lines correspond to analytical solutions from Ref. [10, see Eq. (11)], symbols correspond to simulation results: open diamonds represent simulations where the exchange and dipolar interactions are taken into account and filled circles represent simulations for the case when the dipolar interaction is replaced by the single-ion anisotropy with the coefficient $K_m$, see (S11). (b) Homogeneous equilibrium state in AFM helices with different curvature and torsion; solid lines correspond to analytical solutions (S23), filled circles represent simulations for the case when the dipolar interaction is replaced by the single-ion anisotropy with the coefficient $K$, see (S15). Other notations are similar to (a). Red dashed lines in (a) and (b) correspond to the boundary between the homogeneous (white background) and periodic (gray background) states. (c) Diagram of equilibrium states for an AFM helix spin chain. Open symbols and triangles correspond to periodic and homogeneous states, respectively, obtained in spin-lattice simulations. Solid red curve $\sigma_b(\varkappa)$ shows the boundary between the states. Green dashed line corresponds to the states boundary asymptotic $\sigma_b \approx 0.85\varkappa$ for $\varkappa \ll 1$. The black curve $\sigma_c(\varkappa)$ describes the boundary of linear instability of the homogeneous state, see (S47).

### A. Ferromagnetic ordering

If magnetic moments are coupled ferromagnetically within the spin chain, the second term in (S9) reads

$$\mathscr{H}_{\mathrm{D}}^{\mathrm{AN}} = -\frac{3\mu^2}{a^3} \sum_{\substack{i=-\infty \\ j=1}}^{\infty} \frac{m_{i,x} m_{i+j,x}}{j^3}, \tag{S10}$$

where $m_{i,x}$ is the projection of $i$-th unit magnetic moment on $\boldsymbol{e}_x$. The continuum counterpart of Eq. (S10) for a homogeneous state $\boldsymbol{m}_i = \mathrm{const}$ reads

$$\mathscr{E}_{\mathrm{A}}^{\mathrm{D}} = -K_m \int_{-\infty}^{\infty} (\boldsymbol{m} \cdot \boldsymbol{e}_x)^2 \mathrm{d}x, \qquad K_m = \frac{3\zeta(3)\mu^2}{a^4} \approx 3.6 \frac{\mu^2}{a^4}, \tag{S11}$$

where $\boldsymbol{m}$ is the unit vector of the magnetization and $K_m$ is the effective anisotropy constant with $\zeta(\bullet)$ being the Riemann zeta function [7], see Fig. S1(a) for comparison with simulations.

It is instructive to compare the effective anisotropy (S11) with the thin wire limit. The nonlocal magnetostatic interaction for thin wires of circular and square cross sections is known to be reduced to the local effective easy-tangential anisotropy, even for the case of curved wires [8]. The magnetostatically induced effective anisotropy reads $K^{\mathrm{ms}} = \pi M_{\mathrm{s}}^2$ with $M_{\mathrm{s}}$ being the saturation magnetization of the ferromagnet. The numerical value of dimensionless anisotropy constant $K_m a^4/\mu^2 = 3\zeta(3) \approx 3.6$ is close enough to its continuum counterpart $K^{\mathrm{ms}}/M_{\mathrm{s}}^2 = \pi$. Therefore, we can conclude that the pure 1D dipolar interaction induces the effective anisotropy, which mimics the anisotropy of thin wire [9, page 869]. Nevertheless the difference of about 13% is the main reason for the deviation of previous analytical predictions from the numerical simulations data, see [10].

### B. Antiferromagnetic ordering

To derive a continuum counterpart of (S9) for the case of the antiferromagnetic spin ordering, it is convenient to introduce magnetic moments $\boldsymbol{m}_i^{\mathrm{I}}$ and $\boldsymbol{m}_i^{\mathrm{II}}$ for each of the two sublattices. Indices "I" and "II" label odd and even

magnetic moments, respectively. In terms of the magnetic moments of the sublattices, the anisotropic part of Eq. (S9) reads

$$\mathscr{H}_\text{D}^\text{AN} = -\frac{3\mu^2}{2a^3} \sum_{i=-\infty}^{\infty} \sum_{j=1}^{\infty} \left( \frac{m_{2i,x}^\text{I} m_{2(i+j),x}^\text{I}}{8j^3} + \frac{m_{2i,x}^\text{I} m_{2(i+j)-1,x}^\text{II}}{(2j-1)^3} + \frac{m_{2i+1,x}^\text{II} m_{2(i+j)+1,x}^\text{II}}{8j^3} + \frac{m_{2i+1,x}^\text{II} m_{2(i+j),x}^\text{I}}{(2j-1)^3} \right). \quad \text{(S12)}$$

For the homogeneous state $\boldsymbol{m}_i^\text{I} = -\boldsymbol{m}_j^\text{II}$ for any $i$ and $j$. Therefore the energy (S12) can be rewritten as follows:

$$\mathscr{H}_\text{D}^\text{AN} = -\frac{K_m a}{16} \sum_{i=-\infty}^{\infty} \left[ \left(m_{2i,x}^\text{I}\right)^2 + \left(m_{2i+1,x}^\text{II}\right)^2 + 14 m_{i,x}^\text{I} m_{i,x}^\text{II} \right]. \quad \text{(S13)}$$

The continuum counterpart of (S13) reads

$$\mathscr{E}_\text{D}^\text{AN} = -\frac{K_m}{16} \int_{-\infty}^{\infty} \left[ \left(m_x^\text{I}\right)^2 + \left(m_x^\text{II}\right)^2 + 14 m_x^\text{I} m_x^\text{II} \right] \text{d}x. \quad \text{(S14)}$$

In terms of the Néel and total magnetization vectors, Eq. (S14) reads

$$\mathscr{E}_\text{D}^\text{AN} = -K_m \int_{-\infty}^{\infty} m_x^2 \text{d}x + K \int_{-\infty}^{\infty} n_x^2 \text{d}x, \qquad K = \frac{3}{4} K_m = \frac{9\zeta(3)\mu^2}{4a^4} \approx 2.7 \frac{\mu^2}{a^4}, \quad \text{(S15)}$$

where $K$ is the effective hard-axis anisotropy constant for the Néel vector. It is instructive to note, while the magnetostatic energy is typically negligible for the AFM, the magnitude of the dipolar induced anisotropy in AFM, see (S15) is comparable with one in FM, see (S11).

### C. Curvilinear antiferromagnetic spin chains

The expression (S15) remains valid for the case of curvilinear chains if the curvature and torsion radii are much larger than the distance between magnetic moments, i.e., $\kappa a, |\tau| a \ll 1$. We numerically verify the limits of applicability by the evaluation of the effective anisotropy coefficients for a spin chain (S3) with $s = ai$ and $i$ enumerating magnetic moments. Then, the anisotropic part of (S12) reads

$$\mathscr{H}_\text{D}^\text{AN} = -\frac{3\mu^2}{2} \sum_{i=-\infty}^{\infty} \sum_{j=1}^{\infty} \left[ \frac{(\boldsymbol{m}_{2i}^\text{I} \cdot \boldsymbol{\gamma}_{2i,2(i+j)})(\boldsymbol{m}_{2(i+j)}^\text{I} \cdot \boldsymbol{\gamma}_{2i,2(i+j)})}{|\boldsymbol{\gamma}_{2i,2(i+j)}|^5} + \frac{(\boldsymbol{m}_{2i+1}^\text{II} \cdot \boldsymbol{\gamma}_{2i+1,2(i+j)+1})(\boldsymbol{m}_{2(i+j)+1}^\text{II} \cdot \boldsymbol{\gamma}_{2i+1,2(i+j)+1})}{|\boldsymbol{\gamma}_{2i+1,2(i+j)+1}|^5} \right.$$

$$\left. + \frac{(\boldsymbol{m}_{2i}^\text{I} \cdot \boldsymbol{\gamma}_{2i,2(i+j)-1})(\boldsymbol{m}_{2(i+j)-1}^\text{II} \cdot \boldsymbol{\gamma}_{2i,2(i+j)-1})}{|\boldsymbol{\gamma}_{2i,2(i+j)-1}|^5} + \frac{(\boldsymbol{m}_{2i+1}^\text{II} \cdot \boldsymbol{\gamma}_{2i+1,2(i+j)})(\boldsymbol{m}_{2(i+j)}^\text{I} \cdot \boldsymbol{\gamma}_{2i+1,2(i+j)})}{|\boldsymbol{\gamma}_{2i+1,2(i+j)}|^5} \right], \quad \text{(S16)}$$

where $\boldsymbol{\gamma}_{i,i+j} = \boldsymbol{\gamma}(i+j) - \boldsymbol{\gamma}(i)$. We consider the homogeneous state $\boldsymbol{m}_i^k = m_{i,\alpha}^k \boldsymbol{e}_{i\alpha}$ (in the local reference frame) with $m_{i,\alpha}^k = m_{j,\alpha}^k = \text{const}$ for any $i, j$ with $k = \text{I}, \text{II}$. Then, Eq. (S16) reads

$$\mathscr{H}_\text{D}^\text{AN} = -\frac{3\mu^2}{2} \sum_{i=-\infty}^{\infty} \sum_{\alpha} \sum_{\beta} \left[ \mathfrak{K}_{\alpha\beta}^{11} m_{i,\alpha}^\text{I} m_{i,\beta}^\text{I} + \mathfrak{K}_{\alpha\beta}^{12} m_{i,\alpha}^\text{I} m_{i,\beta}^\text{II} + \mathfrak{K}_{\alpha\beta}^{22} m_{i,\alpha}^\text{II} m_{i,\beta}^\text{II} + \mathfrak{K}_{\alpha\beta}^{21} m_{i,\alpha}^\text{II} m_{i,\beta}^\text{I} \right]. \quad \text{(S17)}$$

Coefficients of the anisotropy, induced by the dipolar interaction, $\mathfrak{K}_{\alpha\beta}^{\nu\upsilon} = \mathfrak{K}_{\alpha\beta}^{\nu\upsilon}(\kappa, \tau)$ with $\nu, \upsilon = 1, 2$ are determined by the system geometry

$$\mathfrak{K}_{\alpha\beta}^{11} = \sum_{j=1}^{\infty} \frac{\left[\boldsymbol{e}_{2i,\alpha} \cdot \boldsymbol{\gamma}_{2i,2(i+j)}\right] \left[\boldsymbol{e}_{2(i+j),\beta} \cdot \boldsymbol{\gamma}_{2i,2(i+j)}\right]}{|\boldsymbol{\gamma}_{2i,2(i+j)}|^5}, \quad \mathfrak{K}_{\alpha\beta}^{22} = \sum_{j=1}^{\infty} \frac{\left[\boldsymbol{e}_{2i+1,\alpha} \cdot \boldsymbol{\gamma}_{2i+1,2(i+j)+1}\right] \left[\boldsymbol{e}_{2(i+j)+1,\beta} \cdot \boldsymbol{\gamma}_{2i+1,2(i+j)+1}\right]}{|\boldsymbol{\gamma}_{2i+1,2(i+j)+1}|^5},$$

$$\mathfrak{K}_{\alpha\beta}^{12} = \sum_{j=1}^{\infty} \frac{\left[\boldsymbol{e}_{2i,\alpha} \cdot \boldsymbol{\gamma}_{2i,2(i+j)-1}\right] \left[\boldsymbol{e}_{2(i+j)-1,\beta} \cdot \boldsymbol{\gamma}_{2i,2(i+j)-1}\right]}{|\boldsymbol{\gamma}_{2i,2(i+j)-1}|^5}, \quad \mathfrak{K}_{\alpha\beta}^{21} = \sum_{j=1}^{\infty} \frac{\left[\boldsymbol{e}_{2i+1,\alpha} \cdot \boldsymbol{\gamma}_{2i+1,2(i+j)}\right] \left[\boldsymbol{e}_{2(i+j),\beta} \cdot \boldsymbol{\gamma}_{2i+1,2(i+j)}\right]}{|\boldsymbol{\gamma}_{2i+1,2(i+j)}|^5}.$$

(S18)

In the case of a helix geometry, the coefficients (S18) do not depend on the index $i$, but on the curvature $\kappa$ and torsion $\tau$.

The continuum counterpart of Eq. (S17) in terms of the Néel and total magnetization vectors reads

$$\mathscr{E}_{\mathrm{D}}^{\mathrm{AN}} = -\int\limits_{-\infty}^{\infty} \left( K_{\alpha\beta}^{mm} m_\alpha m_\beta + K_{\alpha\beta}^{mn} m_\alpha n_\beta + K_{\alpha\beta}^{nm} n_\alpha m_\beta + K_{\alpha\beta}^{nn} n_\alpha n_\beta \right) \mathrm{d}s, \quad (S19)$$

where the coefficients $K_{\alpha\beta}^{fg}$ ($f,g = n,m$) are defined as follows

$$\begin{aligned} K_{\alpha\beta}^{mm} &= \frac{3}{2}\left(\mathfrak{K}_{\alpha\beta}^{11} + \mathfrak{K}_{\alpha\beta}^{12} + \mathfrak{K}_{\alpha\beta}^{22} + \mathfrak{K}_{\alpha\beta}^{21}\right), \quad K_{\alpha\beta}^{mn} = \frac{3}{2}\left(\mathfrak{K}_{\alpha\beta}^{11} - \mathfrak{K}_{\alpha\beta}^{12} - \mathfrak{K}_{\alpha\beta}^{22} + \mathfrak{K}_{\alpha\beta}^{21}\right), \\ K_{\alpha\beta}^{nm} &= \frac{3}{2}\left(\mathfrak{K}_{\alpha\beta}^{11} + \mathfrak{K}_{\alpha\beta}^{12} - \mathfrak{K}_{\alpha\beta}^{22} - \mathfrak{K}_{\alpha\beta}^{21}\right), \quad K_{\alpha\beta}^{nn} = \frac{3}{2}\left(\mathfrak{K}_{\alpha\beta}^{11} - \mathfrak{K}_{\alpha\beta}^{12} + \mathfrak{K}_{\alpha\beta}^{22} - \mathfrak{K}_{\alpha\beta}^{21}\right). \end{aligned} \quad (S20)$$

The coefficients $K_{\alpha\beta}^{fg}$ depend on the curvature $\kappa$ and torsion $\tau$. Their explicit form can be determined by substitution of $\gamma(s = ia)$ [Eq. (S3)] in the respective Eqs. (S18) with further calculation of sums.

In the case of $|\boldsymbol{m}| \ll |\boldsymbol{n}|$, all terms in Eq. (S19), except the last one, can be neglected. Therefore, one can simplify Eq. (S19) as

$$\mathscr{E}_{\mathrm{D}}^{\mathrm{AN}} = K_{\alpha\beta}^{nn} \int\limits_{-\infty}^{\infty} n_\alpha n_\beta \mathrm{d}s. \quad (S21)$$

In the case of a helix with pitch and radius being much larger than the lattice constant $a$, the coefficient $K_{\mathrm{TT}}^{nn}$ of the hard tangential anisotropy makes the major contribution to the anisotropy energy. This is illustrated by Fig. S2, where the relative strength of other anisotropy terms is shown by color. Therefore, one can simplify Eq. (S21) as follows

$$\mathscr{E}_{\mathrm{D}}^{\mathrm{AN}} = K \int\limits_{-\infty}^{\infty} n_{\mathrm{T}}^2 \mathrm{d}s \quad (S22)$$

for $\kappa a \ll 1$ and $\tau a \ll 1$.

## IV. THE HOMOGENEOUS STATE OF AFM HELIX CHAINS

Equilibrium states can be found by solving static Euler–Lagrange equations

$$F(\theta,\phi) = \frac{\partial \mathscr{E}}{\partial \theta} - \frac{\mathrm{d}}{\mathrm{d}\xi}\frac{\partial \mathscr{E}}{\partial(\partial_\xi \theta)} = 0, \quad G(\theta,\phi) = \frac{\partial \mathscr{E}}{\partial \phi} - \frac{\mathrm{d}}{\mathrm{d}\xi}\frac{\partial \mathscr{E}}{\partial(\partial_\xi \phi)} = 0 \quad (S23)$$

with the energy density $\mathscr{E}$ described by (S8).

Let us specify the system: we consider an antiferromagnetic helix chain (S3) with the constant curvature $\varkappa$ and torsion $\sigma$. Namely the constancy of these parameters provides the possibility of the homogeneous (in the curvilinear reference frame) solution. Using the substitution $\theta(s) = \mathrm{const}$ and $\phi(s) = \mathrm{const}$, we get

$$\begin{aligned} \theta_{\mathrm{hom}} &= \pi/2 - \psi, & \phi_{\mathrm{hom}} &= \pi/2, \\ \psi &= \arctan\frac{\varkappa\sigma}{\mathscr{K}_\psi}, & \mathscr{K}_\psi &= \frac{\varkappa^2 - \sigma^2 + \mathscr{K}_0 + 1}{2}, \quad \mathscr{K}_0 = \sqrt{(1 + \varkappa^2 - \sigma^2)^2 + 4\varkappa^2\sigma^2}. \end{aligned} \quad (S24)$$

In the homogeneous state the Néel vector is tilted from the binormal direction by the angle $\psi$ in the TB plane, see Fig. S1(b). In the limit case of small curvatures and torsion, $\varkappa, |\sigma| \ll 1$, the tilted angle $\psi \approx \varkappa\sigma$, cf. the main text. It is instructive to mention that one can introduce the magnetochirality of the state similar to the FM helix. Formally, there exists also the solution with the opposite magnetochirality: $\theta = \pi/2 + \psi$ and $\phi = -\pi/2$. Nevertheless, being the director, the Néel vector is degenerated with respect to the inversion $\boldsymbol{n} \to -\boldsymbol{n}$, or the same, $(\theta,\phi) \to (\pi - \theta, \phi \pm \pi)$, hence both states with opposite magnetochiralities are equivalent.

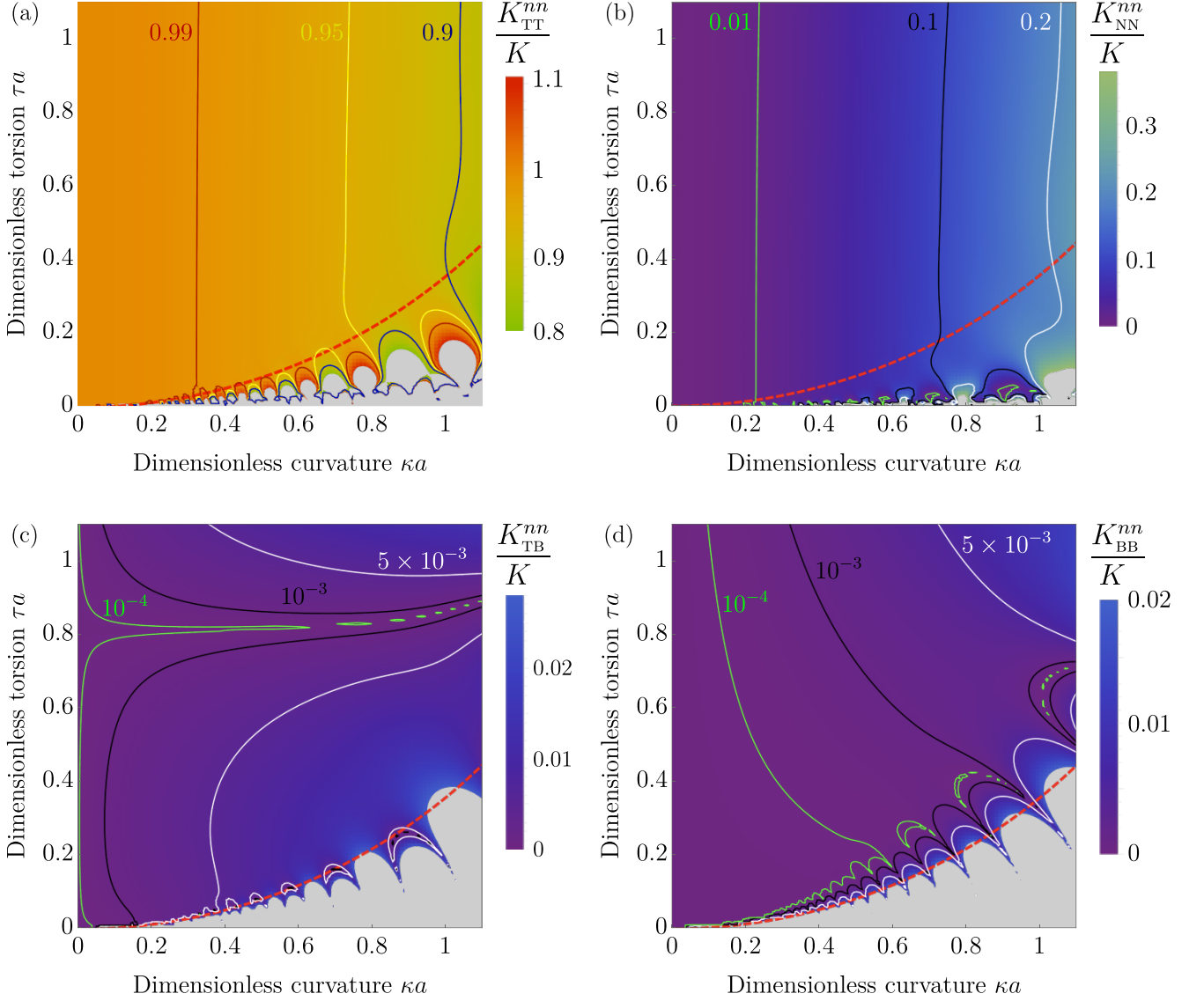

FIG. S2. (Color online) **Diagrams of the coefficients** $K^{nn}_{\alpha\beta}$ (S20), normalized by $K$, for different values of the curvature and torsion. Red dashed line corresponds to the helix pitch $P = 2a$. The coefficients increase exponentially in the gray region. In the region of geometric parameters above the red dashed line the hard-tangential anisotropy with coefficient $K^{nn}_{\mathrm{TT}}$ plays the dominant role in the system.

The homogeneous state takes the simplest form in the rotated reference $\psi$-frame $\{\boldsymbol{e}_1, \boldsymbol{e}_2, \boldsymbol{e}_3\}$. Moreover, the transformation to the $\psi$-frame provides the diagonalization of the tensor of the total anisotropy. By rotating the Néel vector in a local rectifying plane by the angle $\psi$

$$\boldsymbol{n} = U\widetilde{\boldsymbol{n}}, \qquad \widetilde{\boldsymbol{n}} = \{n_1, n_2, n_3\}, \qquad U = \begin{pmatrix} \cos\psi & 0 & \sin\psi \\ 0 & 1 & 0 \\ -\sin\psi & 0 & \cos\psi \end{pmatrix}, \qquad \text{(S25)}$$

one can get the Lagrangian density (S7) in the rotated reference frame:

$$\widetilde{\mathscr{L}} = (\partial_\tau \widetilde{\boldsymbol{n}})^2 - \widetilde{\mathscr{E}}, \qquad \widetilde{\mathscr{E}} = (\partial_\xi n_\iota)(\partial_\xi n_\iota) + \mathscr{K}_1 n_1^2 - \mathscr{K}_3 n_3^2 + \mathscr{D}_1(n_2 \partial_\xi n_3 - n_3 \partial_\xi n_2) + \mathscr{D}_3(n_1 \partial_\xi n_2 - n_2 \partial_\xi n_1) + \widetilde{\mathscr{E}}_0 \quad \text{(S26)}$$

with $n_\iota$ being the components of the Néel vector is the tilted reference frame. The coefficients $\mathscr{K}_1$ and $\mathscr{K}_3$ represent the effective easy-plane and easy-axis anisotropies, respectively, whereas $\mathscr{D}_1$ and $\mathscr{D}_3$ correspond to the effective DMI

coefficients. They read

$$\mathcal{K}_1 = \mathcal{K}_0 - \mathcal{K}_3, \qquad \mathcal{K}_3 = \frac{\varkappa^2 + \sigma^2 + \mathcal{K}_0 - 1}{2}, \qquad \mathcal{D}_1 = 2\sigma \frac{\mathcal{K}_\psi - \varkappa^2}{\sqrt{\mathcal{K}_\psi^2 + \varkappa^2\sigma^2}}, \qquad \mathcal{D}_3 = 2\varkappa \frac{\mathcal{K}_\psi + \sigma^2}{\sqrt{\mathcal{K}_\psi^2 + \varkappa^2\sigma^2}}, \qquad \text{(S27)}$$

with $\widetilde{\mathcal{E}}_0 = \varkappa^2 + \sigma^2$. In the limiting case of small curvatures and torsions

$$\mathcal{K}_0 \approx \mathcal{K}_\psi \approx 1 + \varkappa^2 - \sigma^2, \quad \mathcal{K}_1 \approx 1 - \sigma^2, \quad \mathcal{K}_3 \approx \varkappa^2, \quad \mathcal{D}_1 \approx 2\sigma, \quad \mathcal{D}_3 \approx 2\varkappa, \quad \text{when } \varkappa, |\sigma| \ll 1. \qquad \text{(S28)}$$

Using the angular parametrization for the Néel vector $\widetilde{\boldsymbol{n}} = \boldsymbol{e}_1 \cos\Theta + \boldsymbol{e}_2 \sin\Theta \cos\Phi + \boldsymbol{e}_3 \sin\Theta \sin\Phi$, one can rewrite the Lagrangian (S26) as follows:

$$\widetilde{\mathcal{L}} = (\partial_\tau \Theta)^2 + \sin^2\Theta(\partial_\tau \Phi)^2 - \widetilde{\mathcal{E}}, \qquad \text{(S29a)}$$

$$\widetilde{\mathcal{E}} = (\partial_\xi \Theta)^2 + \sin^2\Theta(\partial_\xi \Phi)^2 + \mathcal{K}_1 \cos^2\Theta - \mathcal{K}_3 \sin^2\Theta \sin^2\Phi + \mathcal{D}_1 \sin^2\Theta \partial_\xi \Phi + 2\mathcal{D}_3 \sin^2\Theta \cos\Phi \partial_\xi \Theta + \widetilde{\mathcal{E}}_0. \qquad \text{(S29b)}$$

The homogeneous solution (S24) in the $\psi$-frame reads $\Theta_\text{hom} = \pi/2$ and $\Phi_\text{hom} = \pi/2$. Its energy

$$\mathcal{E}_\text{hom} = \widetilde{\mathcal{E}}_0 - \mathcal{K}_3. \qquad \text{(S30)}$$

## V. THE PERIODIC STATE OF AFM HELIX CHAINS

The homogeneous state can become unfavorable with increasing curvature and torsion. Taking into account the symmetry of the helix geometry, we are looking for the periodic state

$$\theta_\text{per} = \frac{\pi}{2} + \bar{\theta}(\chi), \qquad \phi_\text{per} = -\chi + \bar{\phi}(\chi) \qquad \text{(S31a)}$$

with $\bar{\theta}(\chi)$ and $\bar{\phi}(\chi)$ being $2\pi$-periodic functions of the angular variable $\chi = \sqrt{\varkappa^2 + \sigma^2}\xi$. The symmetry of the static equations (S23) dictates the symmetry of $2\pi$-periodic functions $\bar{\theta}(\chi)$ and $\bar{\phi}(\chi)$, which have the following Fourier expansion

$$\bar{\theta}(\chi) = \sum_{\eta=1}^{\mathfrak{N}} \epsilon^\eta A_\eta \sin(2\eta - 1)\chi, \qquad \bar{\phi}(\chi) = \sum_{\eta=1}^{\mathfrak{N}} \epsilon^\eta B_\eta \sin 2\eta\chi, \qquad \text{(S31b)}$$

where $\mathfrak{N} \to \infty$.

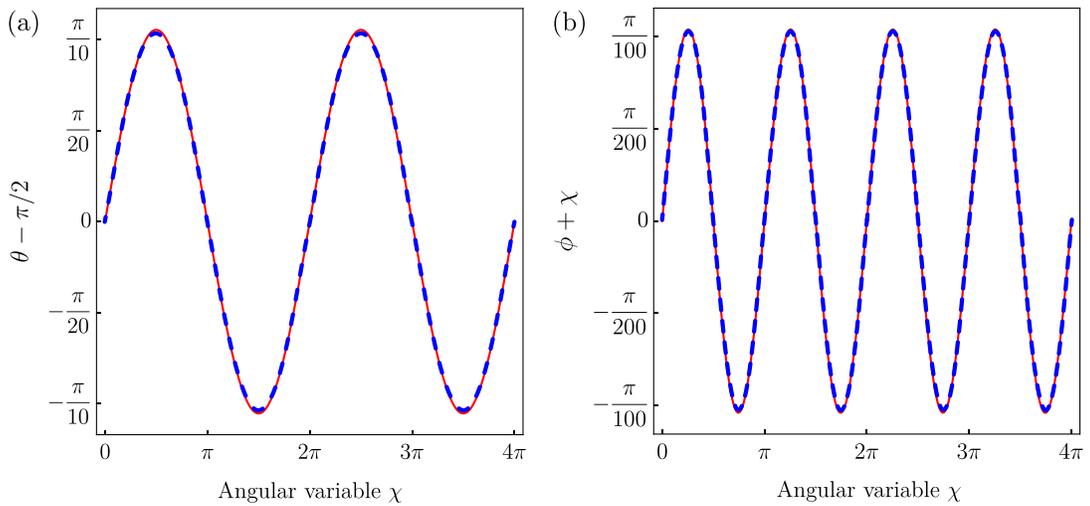

FIG. S3. (Color online) **Periodic state in AFM helix**: magnetic angles (a) $\theta$ and (b) $\phi$ as functions of the angular variable $\chi$ along the helix. Solid lines correspond to the numerical solution of Eqs. (S23) for $\varkappa = 0.7$ and $\sigma = 1.4$. Dashed lines correspond to the solution (S31) with $\mathfrak{N} = 3$.

Numerically, we substitute (S31) into the static equations (S23) and expand results over $\epsilon$ up to the $\mathfrak{N}$-th order:

$$F(\theta, \phi) = \sum_{\eta=1}^{\mathfrak{N}} F_\eta(A_1, \ldots, A_\eta; B_1, \ldots, B_\eta) \sin(2\eta - 1)\chi,$$
$$G(\theta, \phi) = \sum_{\eta=1}^{\mathfrak{N}} G_\eta(A_1, \ldots, A_\eta; B_1, \ldots, B_\eta) \sin 2\eta\chi. \quad (S32)$$

The unknown amplitudes $A_i$ and $B_i$ are found numerically as the solutions of nonlinear polynomial equations

$$F_\eta(A_1, \ldots, A_n; B_1, \ldots, B_\eta) = 0, \quad G_\eta(A_1, \ldots, A_n; B_1, \ldots, B_\eta) = 0, \quad \eta = \overline{1, \mathfrak{N}}, \quad (S33)$$

To calculate the energy of the periodic state $\mathscr{E}_{\text{per}}$, we substitute (S31) into the energy (S8), expand the results over $\epsilon$ up to the $2\mathfrak{N}$-th order and average the energy over the helix period,

$$\mathscr{E}_{\text{per}}(\sigma, \varkappa) = \frac{1}{2\pi} \int_0^{2\pi} \mathscr{E}(A_1, \ldots, A_\eta; B_1, \ldots, B_\eta) \mathrm{d}\chi. \quad (S34)$$

## VI. THE BOUNDARY BETWEEN STATES

The boundary between homogeneous and periodic states $\sigma_b(\varkappa)$, illustrated in Fig. S1(c) [and Fig.1(c) of the main text] are determined numerically, using $\mathfrak{N} = 3$ in (S31). Further increase of $\mathfrak{N}$ does not make significant adjustment to the amplitudes $A_\eta$, $B_\eta$.

It is instructive to obtain the asymptotic behavior of the boundary $\sigma_b(\varkappa)$ in the case $\varkappa, |\sigma| \ll 1$. In this approximation, the homogeneous state reads

$$\theta_{\text{hom}} \approx \frac{\pi}{2} - \varkappa\sigma, \quad \phi_{\text{hom}} = \frac{\pi}{2}, \quad (S35)$$

whereas the periodic one is determined as follows

$$\theta_{\text{per}} = \frac{\pi}{2}, \quad \phi_{\text{per}} \approx -\sqrt{\varkappa^2 + \sigma^2}\xi. \quad (S36)$$

We substitute (S35) in (S8) and expand it into series with respect to $\varepsilon$. It is assumed, that $\varkappa = \varepsilon\varkappa_\epsilon$ and $\sigma = \varepsilon\sigma_\epsilon$. Then, the energy of the homogeneous state per one helix period reads

$$\mathscr{E}_{\text{hom}} = \sigma^2 + \mathcal{O}\left(\varepsilon^3\right). \quad (S37)$$

The average energy of the periodic state per one helix period can be obtained with the same procedure. It reads

$$\mathscr{E}_{\text{per}} = \frac{3}{2}\varkappa^2 + 2\sigma\left(\sigma - \sqrt{\varkappa^2 + \sigma^2}\right) + \mathcal{O}\left(\varepsilon^3\right). \quad (S38)$$

The boundary between states corresponds to the condition $\mathscr{E}_{\text{hom}} = \mathscr{E}_{\text{per}}$. It gives the asymptotic relation between curvature and torsion as

$$\sigma_b(\varkappa) = \sigma_0\varkappa, \quad \sigma_0 = \sqrt{\frac{2\sqrt{7} - 1}{6}} \approx 0.85. \quad (S39)$$

The numerically obtained boundary between states in Fig. 1 of the main text can be fitted as

$$\sigma_b^{\text{trial}}(\varkappa) = \sum_{i=0}^{2} \sigma_i \varkappa^{i+1}, \quad (S40)$$

where $\sigma_1 \approx -0.12$ and $\sigma_2 \approx 0.03$. The function $\sigma_b^{\text{trial}}(\varkappa)$ fits the numerically calculated curve with the accuracy of about $1 \times 10^{-5}$, also see Fig. S1(c).

## VII. THE GROUND STATE OF AFM FLAT CHAINS

Flat curves are characterized by $\tau = 0$ (the torsion-related DMI $d_\text{T}$ is absent) and curvature $\kappa$ with alternating sign in Eq. (S8). In this case, the easy axis $\boldsymbol{e}_3$ is along the $\boldsymbol{e}_\text{B}$, which is oriented perpendicular to the plane of the flat curve. The hard axis $\boldsymbol{e}_1$ is tangential to the curve and is oriented along $\boldsymbol{e}_\text{T}$. The energy density $\mathscr{E}$ in the Eq. (S8) has a positively defined quadratic form. It gives the only ground state $\mathscr{E}_0 = 0$ with $\theta_0 = \phi_0 = \pi/2$, which is analogous of the homogeneous state discussed for the case of the helix. The periodic state is absent in the case of flat AFM spin chains because the spiraling direction of the curvature-induced DMI $d_\text{B}$ coincides with the hard axis $\boldsymbol{e}_\text{T}$ of the dipole-induced anisotropy.

The curvature-induced easy axis anisotropy oriented perpendicular to the plane can contribute to the model Hamiltonian, which is used for the description of Cr-based molecular wheels with nearest-neighbor exchange and easy axis perpendicular to the wheel plane [11, 12].

## VIII. SPIN WAVES IN AFM FLAT CHAINS

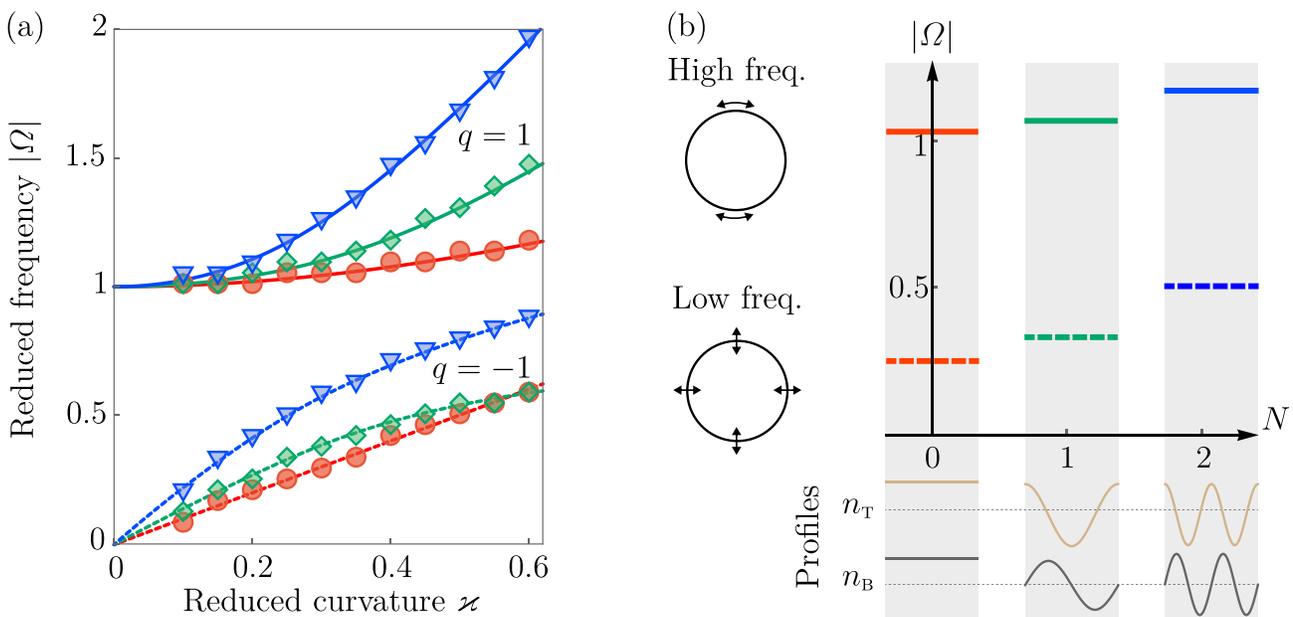

FIG. S4. (Color online) **Spin-wave spectrum of the AFM ring:** (a) The lowest eigenfrequencies of linear excitations as a function of the curvature. Solid and dashed lines correspond to the case of high frequency levels with $q = 1$ and low frequency ones with $q = -1$ in (S43), respectively. Symbols represent the simulation results. (b) Standing waves for different quantum numbers $N$. Schematics shows the oscillation type for the given frequency level. Inset below shows profiles of components of the Néel vector $\boldsymbol{n}$ on the T and B axes.

To illustrate the linear excitations in a flat AFM ring with the ground state $\boldsymbol{n} = \boldsymbol{e}_\text{B}$, we linearize the Euler–Lagrange equations for the Lagrangian (S8) by $\theta(\xi, t) = \pi/2 + \vartheta(\xi, t)$ and $\phi(\xi, t) = \pi/2 + \varphi(\xi, t)$. Here, $\vartheta(\xi, t)$ and $\varphi(\xi, t)$ are small deviations from the equilibrium state. The corresponding equations read

$$\partial_{\xi\xi}\vartheta - \partial_{\tau\tau}\vartheta = \mathscr{K}_0 \vartheta + \mathscr{D}_3 \partial_\xi \varphi, \qquad \partial_{\xi\xi}\varphi - \partial_{\tau\tau}\varphi = \mathscr{K}_3 \varphi - \mathscr{D}_3 \partial_\xi \vartheta. \tag{S41}$$

It is instructive to compare Eqs. (S41) with their dimensional form

$$A\vartheta'' - Ac^{-2}\ddot{\vartheta} = K_0 \vartheta + D_3 \varphi', \qquad A\varphi'' - Ac^{-2}\ddot{\varphi} = K_3 \varphi - D_3 \vartheta', \tag{4}$$

The constants in the (4) can be obtained by the scaling $K_{0,3} = A\ell^{-2}\mathscr{K}_{0,3}$ and $D_3 = A\ell^{-1}\mathscr{D}_3$. The dispersion law for magnons can be written using the plane waves ansatz

$$\vartheta(\xi, \tau) = \vartheta_N \cos(\varkappa N - \Omega \tau), \qquad \varphi(\xi, \tau) = \varphi_N \cos(\varkappa N - \Omega \tau), \tag{S42}$$

where $\vartheta_N$, $\varphi_N$ are small amplitudes, $N \in \mathbb{Z}_+$ is the quantum number and $\varOmega = \Omega/\omega_0$ is the reduced frequency. Then, the dispersion reads

$$\varOmega^2 = \frac{\mathcal{K}_0 + \mathcal{K}_3}{2} + \varkappa^2 N^2 + \frac{q}{2}\sqrt{\mathcal{K}_1^2 + 4\mathcal{D}_3^2 \varkappa^2 N^2}. \tag{S43}$$

The high frequency levels with $q = 1$ take nonzero values for any curvature whereas the low frequency ones with $q = -1$ are strongly influenced by the curvature. Low and high frequency levels accommodate modes, which are oscillations in normal and rectifying surfaces, analogous to $zx$ and $zy$ modes in straight systems [13]), see Fig. S4(b). The quantum number $N$ determines the type of oscillations of the Néel vector, i.e. $N = 0$ corresponds to the homogeneous oscillation, $N = 1$ gives the uniform rotation etc.

## IX. SPIN WAVES IN AFM HELICES AND STABILITY OF THE HOMOGENEOUS STATE

To describe spin waves in an AFM helix, we linearize dynamic equations, which follow from the Lagrangian (S29), on the background of the homogeneous state, $\Theta(\xi,\tau) = \pi/2 + \vartheta(\xi,\tau)$ and $\Phi(\xi,\tau) = \pi/2 + \varphi(\xi,\tau)$. Similar to the ring case, we get the linear set of equations in the form (S41). The dispersion law for spin waves is analyzed by considering the solutions of the linear set of equations (S41) in the form of plane waves,

$$\vartheta(\xi,\tau) = \vartheta_\mathcal{k} \cos(\mathcal{k}\xi - \varOmega\tau), \qquad \varphi(\xi,\tau) = \varphi_\mathcal{k} \sin(\mathcal{k}\xi - \varOmega\tau), \tag{S44}$$

where $\vartheta_\mathcal{k}, \varphi_\mathcal{k} \in \mathbb{R}$ are constant spin-wave amplitudes. The wave vector $\mathcal{k} = \mathcal{k}\boldsymbol{e}_\mathrm{T}$ is oriented along the tangential direction, $\mathcal{k} = k\ell$ is the reduced wave number. By substituting the plane wave solutions (S44) into the linearized equations (S41), we get the dispersion law for the spin waves:

$$\varOmega^2 = \frac{\mathcal{K}_0 + \mathcal{K}_3}{2} + \mathcal{k}^2 + \frac{q}{2}\sqrt{\mathcal{K}_1^2 + 4\mathcal{D}_3^2 \mathcal{k}^2}, \tag{S45}$$

cf. Eq. (5) of the main text. The spectrum consists of two branches with $q = \pm 1$. The low frequency branch with $q = -1$ has a gap $\delta$. The depth of the gap increases with curvature and torsion, see Fig. 2(c) of the main text. There is a critical value of torsion $\sigma_c(\varkappa)$, where the gap vanishes, which results in the instability of the homogeneous state and is determined by the conditions

$$\varOmega(\mathcal{k}_c)\Big|_{q=-1} = 0, \qquad \partial_\mathcal{k} \varOmega(\mathcal{k}_c)\Big|_{q=-1} = 0 \tag{S46}$$

with the critical value of the wave number $\mathcal{k}_c = \sqrt{\mathcal{D}_3^4 - \mathcal{K}_1^2}/(2\mathcal{D}_3)$. The critical curve $\sigma_c(\varkappa)$ can be found as a solution of the algebraic equation:

$$\mathcal{D}_3^2 \left[2\left(\mathcal{K}_0 + \mathcal{K}_3\right) - \mathcal{D}_3^2\right] = \mathcal{K}_1^2 \tag{S47}$$

The numerically calculated critical curve is plotted with a black line in Fig. S1(c). In the region between the boundary curve $\sigma_b(\varkappa)$ and the instability curve $\sigma_c(\varkappa)$, see Fig. S1(c), the homogeneous state becomes metastable.

## X. SIMULATIONS

We numerically investigate static and dynamic states in AFM helices and rings using the spin lattice simulator SLaSi [14, 15]. SLaSi supports arbitrary chain geometries and periodic boundary conditions. For the numerical investigations we consider a classical chain of magnetic moments $\boldsymbol{m}_i$, $i = \overline{1,\mathcal{N}}$, placed at the respective geometry (helix or ring) with $\mathcal{N}$ being the number of sites in the chain. The dynamics of each magnetic moment is governed by the Landau–Lifshitz–Gilbert equation for the Hamiltonian

$$\mathcal{H} = -\frac{JS^2}{2}\sum_i \boldsymbol{m}_i \cdot \boldsymbol{m}_{i+1} - \frac{\mu}{2}\sum_i \boldsymbol{m}_i \cdot \boldsymbol{H}_i^d. \tag{1}$$

Here, $S$ is the spin length, $J < 0$ is the exchange integral, $\mu = g\mu_\mathrm{B}S$ is the total magnetic moment of one site with $g$ being Landé factor, and $\mu_\mathrm{B}$ being Bohr magneton. The dipolar field at $i$-th site reads

$$\boldsymbol{H}_i^d = -\mu \sum_{j=1+i}^{\infty} \frac{\boldsymbol{m}_j r_{ij}^2 - 3\boldsymbol{r}_{ij}(\boldsymbol{m}_j \cdot \boldsymbol{r}_{ij})}{r_{ij}^5} \tag{S48}$$

with $\boldsymbol{r}_{ij}$ being the radius-vector between $i$-th and $j$-th sites and the distance between neighboring sites ia equal $a$.

To analyze the static states, the overdamped system (Gilbert damping $\alpha_\text{G} = 0.5$) with the magnetic length $\ell = 10a$ is considered. The ground state for a given helix geometry is determined as a state of the minimal energy obtained via the relaxation dynamics from (i) three homogeneous states for the Néel vector along axes of the Cartesian reference frame; (ii) the homogeneous state in the local reference frame; (iii) random distribution of magnetic moments.

To verify the accuracy of the effective anisotropy approach, we performed simulations with the dipolar interaction replaced by a signle-ion anisotropy $\mathscr{H}_\text{a} = K_\text{a} \sum_i (\boldsymbol{\mu}_i \cdot \boldsymbol{e}_{\text{T}i})^2$, where $K_\text{a}$ is the anisotropy coefficient and $\boldsymbol{e}_{\text{T}i} = \boldsymbol{e}_\text{T}$ ($s = ai$). The $K_\text{a}$ is chosen to match the resulting magnetic length $\ell$ with the case of the dipolar interaction, see Fig. S1(b).

To obtain the dispersion of spin waves with respect to the ring curvature, we perform a set of simulations for AFM rings of equal length with $\mathscr{N} = 500$ sites and different magnetic length, which determines the reduced curvature $\varkappa$. The initial magnetic texture is given by (S42) for $N = 0$, $N = 1$ and $N = 2$. The spectrum is obtained numerically from the relaxation dynamics with $\alpha_\text{G} = 0.0001$.

There are two steps to compute magnon spectra in helices, shown in Figs. 2 (a, c) of the main text: During the first step, the system is relaxed in the overdamped regime ($\alpha_\text{G} = 0.5$) in the external magnetic field

$$\boldsymbol{H}_\text{EXT}(i) = (-1)^i H_0 \text{sinc}\left(\frac{i - \mathscr{N}/2}{\lambda}\right) \boldsymbol{e}_{\text{T}i}, \tag{S49}$$

where $\text{sinc}(x) = \sin(x)/x$ [16], and $H_0 = 1$ mT is the amplitude and $\lambda = \ell/(2a)$ determines the maximal excited wave number. The initial distribution of magnetic moments was aligned parallel or antiparallel to the binormal. In the second step, the field is switched off, and the magnetization dynamics is initiated under the nominal damping parameter $\alpha_\text{G} = 0.0001$. Then the spatial-temporal Fourier transform is performed for one of the magnetization components to study the spin wave spectrum.

We also perform similar simulations of helices with the FM exchange constant ($J > 0$). We use the value of $K_m$ as an effective anisotropy constant for systems with magnetic dipole-dipole interaction to satisfy the chosen magnetic length by appropriately choosing $J$. The simulations and analysis were made in the same way as for the case of AFM helices. The results are presented in Fig. S1(a), (c.f. Fig. 2 in Ref. 10).

Simulations were performed using the High–performance computing clusters of Taras Shevchenko National University of Kyiv [17] and IFW Dresden.



\end{document}